\documentclass[%
 reprint,
 superscriptaddress,
 amsmath,amssymb,
 aps,
]{revtex4-2}

\usepackage{graphicx}% Include figure files
\usepackage{dcolumn}% Align table columns on decimal point
\usepackage{bm}% bold math
\usepackage{amsmath}
\usepackage{endnotes}
\usepackage{braket}
\usepackage{upgreek}
\usepackage{hyperref} 
\usepackage{ulem}
\usepackage{color}

\begin{document}

\preprint{APS/123-QED}

\title{Quantum Tidal Locking in Orbiting Bose-Einstein Condensates}

%\textcolor{blue}{Quantum Tidal Locking and Emergent Vortex Arrays in Orbiting Bose-Einstein Condensates}% Force line breaks with \\
%%%%%%%%%%%

\author{Yaoyuan Fan}
\thanks{These authors contributed equally to this work}
\affiliation{State Key Laboratory of Advanced Optical Communication Systems and Networks, School of Electronics, Peking University, Beijing 100871, China}
\author{Shuoyu Shi}
\thanks{These authors contributed equally to this work}
\affiliation{School of Physics, Peking University, Beijing 100871, China}
\author{Lang Cao}
\affiliation{State Key Laboratory of Advanced Optical Communication Systems and Networks, School of Electronics, Peking University, Beijing 100871, China}
\author{Ziyue He}
\affiliation{State Key Laboratory of Advanced Optical Communication Systems and Networks, School of Electronics, Peking University, Beijing 100871, China}
\author{Qiuxin Zhang}
\affiliation{National Key Laboratory of Metrology and Calibration, Beijing Changcheng Institute of Metrology \& Measurement, Beijing 100095,China}
\author{Dong Hu}
\affiliation{National Key Laboratory of Metrology and Calibration, Beijing Changcheng Institute of Metrology \& Measurement, Beijing 100095,China}
\author{\\Yu Wang}
\affiliation{National Key Laboratory of Metrology and Calibration, Beijing Changcheng Institute of Metrology \& Measurement, Beijing 100095,China}
\author{Qing Wang}
\email{qw1988@pku.edu.cn}
\affiliation{State Key Laboratory of Advanced Optical Communication Systems and Networks, School of Electronics, Peking University, Beijing 100871, China}
\author{Tianwei Zhou}
\email{zhou@lens.unifi.it}
\affiliation{Department of Physics and Astronomy, University of Florence, 50019 Sesto Fiorentino, Italy}
\author{Xiaoji Zhou}
\affiliation{State Key Laboratory of Advanced Optical Communication Systems and Networks, School of Electronics, Peking University, Beijing 100871, China}
\affiliation{Institute of Carbon-Based Thin Film Electronics, Peking University, Shanxi, Taiyuan 030012, China}


\begin{abstract}

Angular momentum coupling manifests widely in diverse physical systems, underpinning the emergent properties and collective dynamics across different scales. The tidal locking, which originates from the synchronization of rotational and orbital motions, has far-reaching impacts in celestial mechanics, reflecting fundamental processes of angular momentum transfer, energy dissipation, and evolution toward dynamical equilibrium. However, its counterpart in mesoscopic quantum fluids has remained largely unexplored. Here we demonstrate the emergence of quantum tidal locking in Bose-Einstein condensates undergoing central force motion in an anharmonic  potential. The condensate follows a well-defined orbital trajectory in a static trap and experiences an effective rotating potential induced by the trap anharmonicity. The sustained geometric squeezing continuously deforms the condensate and drives a self-organized synchronization process, in which the intrinsic rotation gradually locks to the orbital motion. Numerical simulations further reveal the formation of a ring-shaped vortex array over longer timescales, {arising from the coherent evolution of the rotating matter wave during the locking dynamics}. Our findings establish quantum tidal locking in mesoscopic systems as a robust self-organized mechanism for generating and stabilizing circulating states.

\end{abstract}

\maketitle

\section{\label{sec1}Introduction}
Angular momentum coupling between rotational and orbital degrees of freedom appears across diverse physical scales. In microscopic regimes, spin-orbit coupling governs fundamental phenomena from atomic fine structure to topological insulators~\cite{RevModPhys.82.3045,Galitski2013,Zhai_2015}. Macroscopically, gravitational interactions drive celestial synchronization through torque-based coupling. These phenomena share a common mathematical essence: their system Hamiltonians contain coupling terms that link intrinsic and orbital angular momenta.

Synchronization of rotational and orbital motions finds a spectacular manifestation in classical tidal locking~\cite{Goldreich1966}, where gravitational torques synchronize the satellite's rotational period with its orbital period, as observed in the Moon-Earth system~\cite{Williams2014}. This classical phenomenon naturally motivates the search for its quantum analog: can tidal locking arise in mesoscopic quantum systems? Among such systems, quantum gases have served as a primary platform, where rotational dynamics are typically studied via external potential stirring in Bose-Einstein condensates (BECs)~\cite{PhysRevLett.83.2498,PhysRevLett.84.806,PhysRevLett.84.2056,science.1060182,PhysRevLett.92.040404,PhysRevLett.92.050403,RevModPhys.81.647,PhysRevA.91.013603,Mukherjee2022}, interacting fermions~\cite{Zwierlein2005,Kwon2021,Lunt2024}, and recently in atomic supersolids~\cite{Poli2025}, placing the reference frame in the rotating potential. However, the direct quantum analog of tidal locking, setting the condensate into orbital motion around a static potential center, remains largely unexplored. Such a scenario not only closes a major gap between celestial mechanics and quantum hydrodynamics, but also enables exploration of the intrinsic coupling between collective orbital motion and internal rotational degrees of freedom in quantum fluids.

\begin{figure}[t!]
	\includegraphics[width=\linewidth]{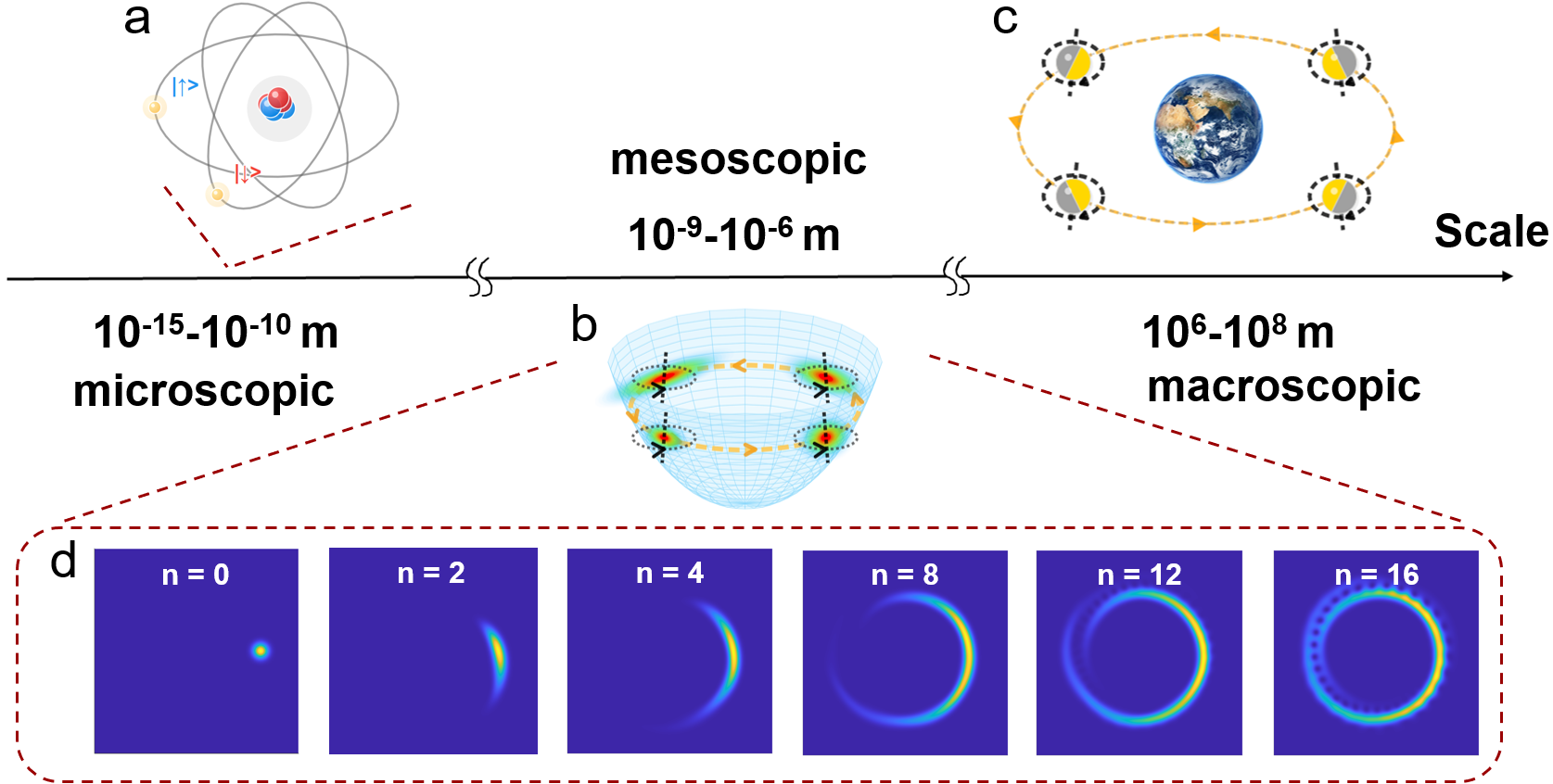} % Here is how to import EPS art
	\caption{\label{fig1} Comparative diagrams of angular momentum coupling mechanisms across physical scales. 
		(a) Microscopic scale: Schematic of electron spin-orbit coupling. 
		(b) Mesoscopic scale: Schematic of rotation-orbit coupling in BECs (quantum tidal locking). 
		(c) Macroscopic scale: Schematic of rotation-orbit coupling in celestial bodies (tidal locking in the Earth-Moon system). 
		(d) Numerical simulation of the atom cloud at different orbital cycles, where $n$ denotes the number of orbital revolutions, demonstrating quantum effects distinct from classical tidal locking.
	}
\end{figure}

Here we report the quantum tidal locking in $^{87}$Rb BECs undergoing orbital motion in a static magnetic quadrupole trap. With a self-consistent combination of theoretical modeling, numerical simulation, and experimental validation, we investigate the rotation-orbit coupling that emerges spontaneously from the breaking of rotational symmetry. Unlike previous studies focusing on externally induced rotation and the resulting vortex nucleation dynamics, the coupling in our system emerges from the orbital motion in a static potential. The condensate follows a well-defined orbit and experiences an effective rotating potential arising from the anharmonicity of the trap. This leads to a synchronization between intrinsic rotation and orbital motion, which can be understood as a direct consequence of the geometric squeezing process. We demonstrate that this coupling drives the system into a synchronized locking state and, on longer timescales, a dynamical formation of ring-shaped vortex arrays (see Fig.~\ref{fig1}), arising from the coherent interference of different parts of the rotating matter wave during the locking dynamics. Beyond its fundamental interest as a quantum analog of a celestial phenomenon, this self-organized synchronization provides a pathway for vortex matter creation.

%%%%%%%%%%%%%%%%%%%%%%%%%%%%%%%%%%%%%%%%%%%%%%%%%%%%%%%%%%
\section{\label{sec2}Theoretical Model}
% In this part, we will derive the rotation-orbit coupling term, and examine its impact on the evolution of intrinsic angular velocity and angular momentum from both analytical and numerical perspectives.
In this section, we show that orbital motion of the center-of-mass (COM) in anharmonic trap will introduce the angular momentum coupling between orbital motion and intrinsic rotation, which creates a body-fixed gauge field. The field forms a effective saddle potential for the guiding center and the wave packet will be stretched according to the geometric squeezing process and gains intrinsic angular momentum, which finally leads to tidal locking phenomena.

To focus on the behavior of BEC under rotation motion, consider a three-dimensional BEC confined in a cylindrical anharmonic trap \(V(x,y,z)=V(\sqrt{x^2+y^2})\), undergoing circular motion of radius \(r_0\) in \(xy\) plane, and its degree of freedom along \(z\) is locked in the ground state of \(V_z=\frac{1}{2}m\omega_z^2z^2\). The 2D Hamiltonian is
\begin{equation}
\label{eq1}
\hat{H} = \frac{\hat{\bm p}^2}{2m} + \frac{m\omega_0^2}{2}\left((1-2\beta)\hat{\bm x}^2+\frac{\beta}{r_0^2}\hat{\bm x}^4\right) + g|\psi|^2,
\end{equation}
where \(\hat{\bm x} = (\hat x,\hat y)\), \(m\) is the atomic mass, \(\omega_0\) the trapping frequency, \(g=4\pi\hbar^2a_s/m\) the interaction strength where \(a_s\) the scattering length~\cite{RevModPhys.71.463}, \(\beta\) the anharmonic parameter that controls the proportion and sign of the quartic term, and \(r_0\) makes \(\beta\) dimensionless.

We first simplify the COM motion, then delve into the intrinsic motion. The COM degree of freedom can be eliminated if we assume that the potential well is slowly varying relative to the scale of wave packet. Therefore, the time evolution of the COM degree of freedom will be approximately independent of the shape of the wave packet, and it will evolve according to classical equations of motion, which is a circular motion around the center of trap with circular frequency \(\Omega\). We only needs to deal with the evolution of wave packet shape while the COM degree of freedom can be regarded as a time-dependent parameter \(\bm{X}(t) = (r_0\cos\Omega t,r_0\sin\Omega t)\).

For intrinsic motion, the slight difference in the local trapping potential in the vicinity of COM is crucial. Due to trap anharmonicity, in different directions \(\bm{l}\), the local trapping frequencies \(\partial ^2V/\partial\bm{l}^2|_\text{com}\) are anisotropic~\cite{PhysRevA.91.013603}. Thus, the local potential experienced by the wave packet is an elliptical harmonic, whose axes are the radial and azimuth, rotating along with COM. For example, when \(\beta < 0\), the trapping frequency maximizes at azimuth, and minimizes at radial. It is the breaking of rotational symmetry that enables the wave packet to experience the rotation of COM.

To describe the intrinsic motion of the wave packet, we consider the Hamiltonian in its body-fix reference frame via canonical transformation, with type-2 generating function
\begin{equation}
    \label{eq2}
    \hat H' = \hat H + \frac{\partial}{\partial t}F_2(\hat{\bm{x}}, \hat{\bm{P}},t).
\end{equation}
First, we transform from laboratory frame \(S_\text{lab}\) to the COM frame \(S_\text{com}\) with non-rotating axes. The generating function and Hamiltonian in \(S_\text{com}\) are 
\begin{align}
    \label{eq3}
    &F_2^{(1)}(\hat{\bm{x}},\hat{\bm{p}}',t) = \left[\hat{\bm{x}}-\bm{X}(t)\right]\cdot\hat{\bm{p}}'+m\dot{\bm{X}}(t)\cdot\hat{\bm{x}}, \\
    \label{eq4}
    &\hat H' = \hat H - \dot{\bm{X}}(t)\cdot\hat{\bm{p}}' + m\ddot{\bm X}(t)\cdot\hat{\bm x}. 
\end{align}
Then, from COM frame \(S_\text{com}\) to the corotating body-fixed frame \(S_\text{body}\), where the axes rotate synchronously with the circular motion of COM [see Fig.~\ref{fig1}(a)], the generating function and Hamiltonian in \(S_\text{body}\) are
\begin{align}
    \label{eq5}
    &F_2^{(2)}(\hat{\bm{x}}',\hat{\bm{p}}'',t) = 
    \begin{pmatrix}
        x' & y'
    \end{pmatrix}
    \begin{pmatrix}
        \cos\Omega t & -\sin\Omega t \\
        \sin\Omega t & \cos\Omega t
    \end{pmatrix}
    \begin{pmatrix}
        p_r \\ p_\theta
    \end{pmatrix}, \\
    \label{eq6}
    &\hat H'' = \hat H' - \Omega(\hat r\hat p_\theta - \hat r_\theta \hat p_r) = \hat H' - \Omega\hat L_{z''}.
\end{align}
Substituting \(\bm{X}(t) = (r_0\cos\Omega t,r_0\sin\Omega t)\) and discarding higher-order terms of \(\hat{\bm{x}}''\) in potential energy, we obtain
\begin{equation}
\label{eq7}
\hat H'' = \frac{\hat p_r^2+\hat p_\theta^2}{2m}+\frac{m}{2}\left(\omega_r^2\hat r^2+\omega_\theta^2 \hat r_\theta^2\right) - \Omega \hat L_{z''} + g|\psi|^2,
\end{equation}
where subscripts \(r\) and \(\theta\) denote radial and azimuth, \(\hat L_{z''} = \hat r\hat p_\theta-\hat r_\theta \hat p_r\) is the angular momentum in \(S_\text{body}\), \(\omega_r\) and \(\omega_\theta\) are the local trapping frequencies along the direction of \(\hat r\) and \(\hat r_\theta\), respectively. The rotation-orbit coupling term \(\hat H_\text{ROC} = -\Omega \hat L_{z''}\)
%\begin{equation} 
%    \label{eq10}
%    \hat H_\text{ROC} = \frac{|\bm{X}\times\dot{\bm{X}}|}{|\bm{X}|^2}\hat L_z,
%\end{equation}
exhibits a structure analogous to atomic spin-orbit coupling, where an effective magnetic field generated by orbital rotation (the COM angular velocity \(\Omega\)) couples to the intrinsic rotation (the intrinsic angular momentum \(\hat L_{z''}\)) or spin. We will show that this rotation-orbit coupling term generates a body-fix gauge field and squeezes the wave packet~\cite{Cooper2008,doi:10.1126/science.aba7202,PhysRevA.105.023310,CRPHYS_2023__24_S3_241_0,chen2025}, and finally leads to the phenomenon of quantum tidal locking.

We derive the Landau-level guiding center and cyclotron as the direct consequence of the rotation-orbit coupling, analyze the resulting noncommutative guiding-center dynamics in the effective potential, and thereby reveal the geometric squeezing process. Equation~\eqref{eq7} can be rewritten
\begin{multline}
    \label{eq8}
    \hat H'' = \frac{1}{2}
    \begin{pmatrix}
        \hat r & \hat p_\theta
    \end{pmatrix}
    \begin{pmatrix}
        m\omega_r^2 & -\Omega\\
        -\Omega & 1/m
    \end{pmatrix}
    \begin{pmatrix}
        \hat r \\ \hat p_\theta
    \end{pmatrix} \\
    + \frac{1}{2}
    \begin{pmatrix}
        \hat r_\theta & \hat p_r
    \end{pmatrix}
    \begin{pmatrix}
        m\omega_\theta^2 & \Omega\\
        \Omega & 1/m
    \end{pmatrix}
    \begin{pmatrix}
        \hat r_\theta \\ \hat p_r
    \end{pmatrix}
     + g|\psi|^2.
\end{multline}
Diagonalization of the above Eq.~\eqref{eq8} yields a new set of coordinates and momenta
\begin{align*}
\label{eq9}
\hat\xi &= \frac{1}{2}\left[\hat r-\frac{\hat p_\theta}{m\omega_r}\right],\;
\hat\eta = \frac{1}{2}\left[\hat r_\theta-\frac{\hat p_r}{m\omega_\theta}\right],\\
\hat R &= \frac{1}{2}\left[\hat r+\frac{\hat p_\theta}{m\omega_r}\right],\,
\hat R_\theta = \frac{1}{2}\left[\hat r_\theta+\frac{\hat p_r}{m\omega_\theta}\right],\tag{9}
\end{align*}
\((\hat\xi,\hat\eta)\) and \((\hat R,\hat R_\theta)\) represent the Landau-level guiding-center and cyclotron motion, respectively. The effective potential of the guiding center is
\begin{equation*}
\label{eq10}
\hat V_\text{gc} = m\omega_r(\omega_r-\Omega)\hat R^2+m\omega_\theta(\omega_\theta-\Omega)\hat R_\theta^2 + V_\text{int}(\hat R,\hat R_\theta),\tag{10}
\end{equation*}
where \(V_\text{int}(\hat R,\hat R_\theta) = g|\psi(\hat R,\hat R_\theta)|^2\) is the interaction term. In circular motion, \(\Omega\) is always between \(\omega_r\) and \(\omega_\theta\) (see Appendix~\hyperref[AppxA]{A}), so \(\hat V_\text{gc}\) is a saddle surface. 

The geometrical squeeze process deforms the wave packet. In the effective saddle potential, guiding centers drift along the isoenergy contours according to \(\bm{v}_d=\bm{\Omega}\times\bm{\nabla} \hat V_\text{gc}/2m\Omega\sqrt{\omega_r\omega_\theta}\) (see Appendix~\hyperref[AppxB]{B}), thus the wave packet will be squeezed along a certain direction and extends along another, which subtends an angle with the azimuth.

The intrinsic angular momentum accumulates through this squeezing process, and the wave packet will eventually rotate synchronously with COM. Initially, the wave packet is translational relative to the COM. Then the guiding centers are squeezed, and the wave packet comes to follow the rotation of \(S_{\text{body}}\), generating nonzero intrinsic angular momentum. As the squeezing becomes significant, the wave packet will act like a rigid body \cite{chen2025}, and the guiding centers will acquire sufficient angular momentum, exhibiting an intrinsic rotation with angular velocity \(\Omega\), and is locked with its COM orbital rotation. This identifies a qualitatively different dynamical regime from previously studied driven condensate systems.

\section{\label{sec3}Numerical Simulation}

To study the locking effect in the long term and the influence of anharmonic parameter \(\beta\) and interacting strength \(g\) in Eq.~\eqref{eq1} more quantitatively, we exploited split-operator method~\cite{doi:10.1137/0723033, BAO2003318, WANG200517} to solve Gross-Pitaevskii equation and simulate the 2D circular motion (detailed in Appendix~\hyperref[AppxC]{C}), as shown in Fig.~\ref{fig23}. Figure~\ref{fig23}(a) shows the density profile from the \(z\) axis. The wave packet extends along the isoenergy contours, which subtends an angle with the azimuth, and then stretch along the azimuthal direction because the angular velocity varies across different orbits in the anharmonic trap. Figure~\ref{fig23}(b) shows the superposition of intrinsic flow and density. The intrinsic flow is observed in the COM frame \(S_\text{com}\), which is calculated after removing the plane-wave component corresponding to the COM velocity \(\exp(\text{i}\bm{k}_\text{com}\cdot\bm{x})\), where \(\hbar\bm{k}_\text{com} = m\bm{v}_\text{com}\). The quadrupolar flow~\cite{PhysRevLett.86.377} is dense in the \uppercase\expandafter{\romannumeral1} and \uppercase\expandafter{\romannumeral3} quadrants and sparse in the rest two quadrants, indicating a nonzero intrinsic angular momentum. We also simulate the situation where the angular momentum quanta is not zero in Appendix~\hyperref[AppxD]{D}, whose results are consistent with Ref.~\cite{CRPHYS_2023__24_S3_241_0}.

\begin{figure}[t!]
\includegraphics[width=\linewidth]{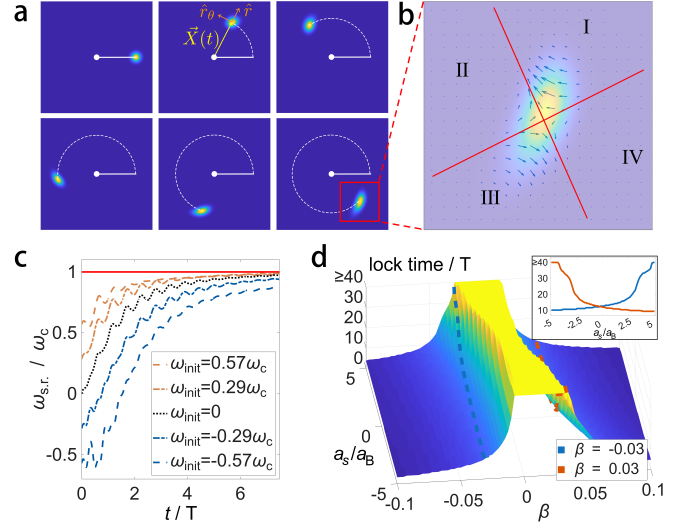} % Here is how to import EPS art
\caption{\label{fig23} 
(a)--(b) Numerical simulation of the density and intrinsic flow fields distributions in orbiting BECs, where \(r_0=|\bm{X}|=90\upmu\text{m}\), \(\omega_0/2\pi=1.25\text{ Hz}\), \(N=2500\) atoms, \(\beta=-0.15\), and \(g\) is calculated with \(a_s=10a_{\rm B}\) ($a_{\rm B}$ is the Bohr radius) in Eq.~\eqref{eq1}: (a) Density profile from the \(z\) axis during orbital motion. (b) Superposition of density and intrinsic flow fields.
(c)--(d) Numerical simulation of angular velocity locking, where \(r_0=|\bm{X}|=80\upmu\text{m}\), \(\omega_0/2\pi = 2.5\text{ Hz}\), and \(N=2500\) atoms: (c) Temporal evolution of intrinsic angular velocity $\omega_\text{i.r.}$ at \(\beta = -0.1\) and \(g = 0\) with different initial values, normalized to the COM angular velocity \(\omega_\text{c}\). (d) The locking time [when \((\omega_\text{c}-\omega_\text{i.r.})/\omega_\text{c} = 1/e\)] under different $\beta$ and $g$, with initial intrinsic angular velocity \(\omega_\text{init}=0\).
}
\end{figure}

\begin{figure*}[t!]
\includegraphics[width=\linewidth]{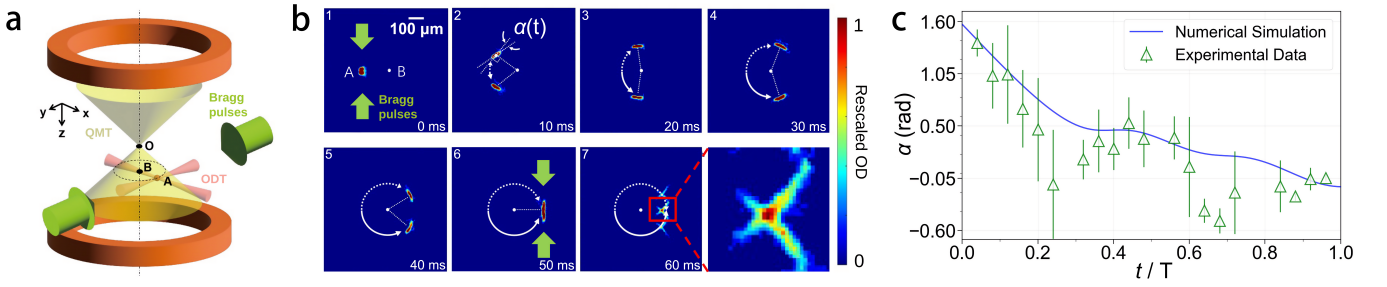} % Here is how to import EPS art
\caption{\label{fig3} Experimental apparatus and results for quantum tidal locking. 
    (a) Schematic of the experimental setup. %: BECs are prepared at the center of the optical dipole trap, marked as point A; then they are transferred in situ to the quadrupole magnetic trap, where the {\color{red}null} point O is adjustable in three dimensions. The Bragg pulse optical lattice is applied to impart initial orbital angular momentum to the BECs. 
    (b) Top-view \textit{in situ} absorption images at different evolution time. Upon excitation by a Bragg pulse at point A, the BECs split into two packets that revolve clockwise and counterclockwise, respectively, around point B in the horizontal plane. The angle between the major axis of each wave packet and the tangent direction of its orbital trajectory is denoted as $\alpha$. Images were captured every 10 ms. At 50 ms (half of the orbital period), a second Bragg pulse was applied. A portion of each atomic cloud was transferred back to the zero‑momentum state and subsequently moved toward point B. The magnified view of this portion is shown in the eighth picture. The X-shaped pattern demonstrates the existence of coupling between the intrinsic spin angular velocity and the orbital angular velocity.
    (c) Temporal evolution of the angle $\alpha$. The error bars represent the standard deviations of five individual measurements. The solid line represents the numerical simulation results, considering the experimental situation for $\beta = -0.05$ and $a_s = 99 a_{\rm B}$.}
\end{figure*}

More quantitative analyzes are shown in Figs.~\ref{fig23}(c)(d). Figure~\ref{fig23}(c) displays the process of angular velocity convergence. The intrinsic angular velocity \(\omega_\text{i.r.}\) (i.r. means intrinsic rotation) is the angular momentum divided by the rigid body inertia momentum~\cite{chen2025}. We imposed a phase field \(\exp[\text{i}k_0(x-X_\text{com})(y-Y_\text{com})]\) on a deformed BEC to simulate the case in Fig.~\ref{fig23}(b), generating a series of initial angular velocity \(\omega_\text{init}\), and discovered that they all converge to the COM angular velocity \(\omega_\text{c}\). Subtle oscillations are caused by the breathing of cyclotron motion~\cite{CRPHYS_2023__24_S3_241_0}. Figure~\ref{fig23}(d) illustrates the influence of anharmonic parameter \(\beta\) and interaction strength \(g\) (tuned by scattering length \(a_s\)~\cite{RevModPhys.71.463}) on the time of angular velocity convergence. The locking time will be much shorter when \(|\beta|\) increases, and diverges when \(|\beta|\rightarrow0\). This is because a larger \(|\beta|\) enlarges the differences between \(\omega_r\), \(\omega_\theta\), and \(\Omega\) in Eq.~\eqref{eq10} and accelerates the drift.

Interestingly, the effect of \(g\) is inverted when the sign of \(\beta\) is reversed (in weak interaction regime); see the inset of Fig.~\ref{fig23}(d). This is because when the wave packet extends, the interaction energy in Eq.~\eqref{eq10} will be changed, and the external potential energy of guiding center must exhibit an opposite change to offset it. For instance, when \(\beta<0\) and \(g>0\)—the conditions in the following experimental section, the extension of wave packet decreases the interaction energy, and the guiding center drifts toward the region of lower external potential energy [azimuthal direction \(\hat{r}_\theta\) in Fig.~\ref{fig1}(a)]. Such drift could broaden or narrow the radial extent of the wave packet, accelerating or decelerating the locking of \(\omega_\text{i.r.}\), respectively.

%%%%%%%%%%%%%%%%%%%%%%%%%%%%%%%%%%%%%%%%%%%%%%%%%%%%%%%%%%

%%%%%%%%%%%%%%%%%%%%%%%%%%%%%%%%%%%%%%%%%%%%%%%%%%%%%%%%%%%
\section{\label{sec4}Experimental Demonstration}

\begin{figure*}[t!]
\includegraphics[width=0.9\linewidth]{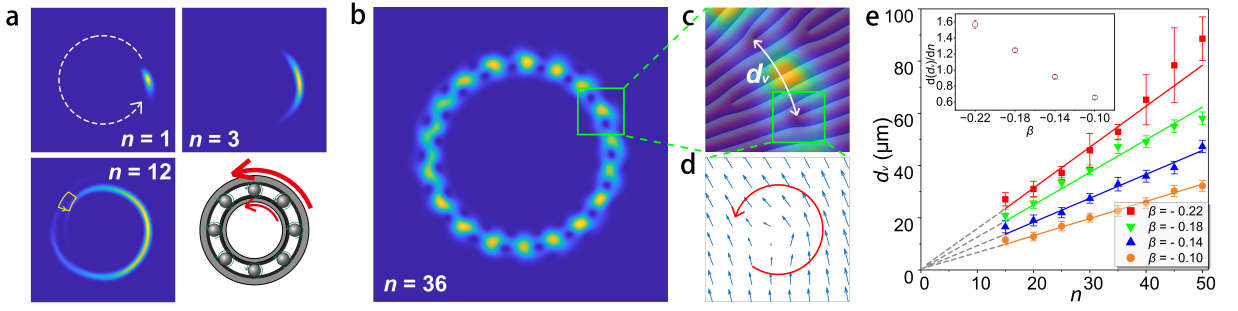} % Here is how to import EPS art
\caption{\label{fig4}
(a) Numerical simulation of the formation of a ring-shaped vortex array (\(r_0=|\bm{X}|=80\upmu\text{m}\), \(\omega_0/2\pi = 2\text{ Hz}\), \(\beta=-0.14\), and \(g=0\)), where \(n=t/\mathrm{T}\) denotes the number of orbital revolutions; and an illustrative diagram of a ball bearing, serving as a visual metaphor. The vector line integral of velocity field over the yellow loop is nonzero, resulting in the vortex arrays.
(b) The ring-shaped vortex array at $n=36$, where vortex are distributed in multiple layers.
(c) Superimposed diagram of density distribution and phase distribution. The feature of singly quantized vortex phase distribution is observed \cite{PhysRevA.102.063330,
PhysRevA.109.043304}.
(d) Local enlargement showing velocity field. Nonzero velocity circulation integral demonstrating the emergence of vortices.
(e) Temporal evolution of the arc distance $d_{\rm v}$ between vortices under different $\beta$. We tracked 8 pairs of neighboring vortices to estimate the average distance. Error bars come from inhomogeneity of the vortex array and measurement errors. The lines are linear fit through the origin. Inset shows that the growth rate of $d_{\rm v}$ varies with anharmonicity \(\beta\).}
\end{figure*}

The experiment starts with a BEC of $2\times 10^4$ $^{87}$Rb atoms at the center [point A in Fig.~\ref{fig3}(a)] of an optical dipole trap~\cite{Yaoyuan}. The condensates, prepared in the $|F=1, m_F=-1\rangle$ hyperfine state, are then transferred \textit{in situ} to the quadrupole magnetic trap (QMT), where the null point O is adjustable in three dimensions. The projection of point O onto the horizontal plane is point B, with BA oriented along the \(x\) axis. By applying a Bragg pulse optical lattice along the \(y\) direction, the initially stationary BEC can be split into two wave packets carrying momentum states of $\pm2\hbar k$ respectively along the \(y\) axis. Through fine tuning of the magnetic field gradient and the position of the magnetic field null point O, the magnetic gradient force on the two atomic wave packets provided the centripetal force required for circular motion in the horizontal direction, while being balanced with gravity in the vertical direction. Under these conditions, the two atomic wave packets undergo circular orbital motion in opposite directions within the horizontal plane. The orbital radius of the atomic clouds is 0.2 mm, and the period for one complete revolution is \(T\) = 100 ms. Figure~\ref{fig3}(b) shows the top-view \textit{in situ} absorption images of the orbiting wave packets, through which we track and extract the evolution information.

The experiment is performed with the potential anharmonicity $\beta = -0.05$, which can be tuned by adjusting the distance between the magnetic field null point O and the orbital center B. Since it is experimentally infeasible to extract the intrinsic angular momentum within the wave packet, we consider $\alpha$ as the observable for characterizing the synchronization of angular momentum, which is the angle between the major axis of the atomic wave packets and the tangential direction. In the classical tidal locking model, under the influence of tidal forces, $\alpha$ oscillates and gradually converges to a steady value. For quantum tidal locking in a BEC, $\alpha$ proves to be a well-suited observable, both in experiments and numerical simulations. 

Figure~\ref{fig3}(c) shows the damped oscillations of experimentally measured $\alpha$ during the orbital motion, exhibiting a solid quantitative agreement with the numerical simulation (blue line, which will converge to zero over longer timescales), with no fitting parameters. It is notable that $\alpha$ initially approaches $\pi$/2 because the influence of the BEC's orbital revolution has not yet manifested, while the lower radial trapping frequency compared to the tangential one causes the wave packet to stretch radially. Over time, the anharmonicity of the potential leads to the emergence of coupling between the orbital angular momentum and the intrinsic angular momentum. As a result, the angle $\alpha$ evolves toward zero through oscillations. Our experimental observations directly demonstrates the atomic wave packet elongation and the generation of intrinsic angular momentum—clear signatures of progression toward the locked state. This phenomenon represents a characteristic feature of the classical tidal locking process, revealing a notable similarity between quantum tidal locking and its classical counterpart.

To further confirm the emergence of intrinsic angular momentum, we applied a second Bragg pulse to quench the orbital motion when the two atomic wave packets met again after completing a half-period of their orbital trajectories. Following 10 ms of evolution in the magnetic trap, the seventh image in Fig.~\ref{fig3}(b) was obtained. The portions where orbital motion was suppressed moved toward point B due to the horizontal magnetic field gradient force. Magnified views reveal the formation of a characteristic X-shaped density distribution, demonstrating that the two wave packets orbiting in opposite directions maintained intrinsic angular momentum with opposite orientations even after their orbital angular momentum was suppressed. This confirms that the generation of intrinsic angular momentum is associated with orbital motion and shares the same direction as the orbital angular momentum.

%%%%%%%%%%%%%%%%%%%%%%%%%%%%%%%%%%%%%%%%%%%%%%%%%%%%%%%%%%%%%%%%%%%%%%
%\subsection{Novel Pathway for Superfluid Vortex Phase Transitions}
%\label{subsec:vortex_pathway }

\section{\label{sec5}Emergent vortex arrays}
% \section{\label{sec5}Discussions and Conclusions}
While our experimental observations confirm the existence of quantum tidal locking within one orbit, numerical simulations reveal that if the BEC could sustain orbital motion for more orbits, a ring-shaped vortex array would appear.

As shown in Fig.~\ref{fig4} we discovered that quantum tidal locking establishes a fundamentally distinct paradigm for generating vortex arrays in Bose-Einstein condensates~\cite{PhysRevLett.92.050403,Kavoulakis_2003, PhysRevA.71.013605}. The vortex formation process through quantum tidal locking evolves through four distinct stages, as shown in Figs.~\ref{fig4}(a)(b). In stage one, at the very beginning, the BEC circulates as a coherent whole around the center of potential with nearly zero intrinsic angular momentum. In stage two, the BEC is locked and begins to stretch along the azimuth due to anharmonic effect; this stage corresponds to our experimental observations. In stage three, the BEC forms a Ouroboros-like shape and begins to form the vortex array. In stage four, the BEC turns into a ring-shaped vortex lattice as predicted by our numerical simulations [Figs.~\ref{fig4}(c)(d)]. This features the degeneracy of rotations where the distinction between orbital and internal rotation vanishes, creating a topologically protected state with equivalent rotational and orbital angular momenta. 

Conventional vortex nucleation methods rely on external potential stirring that excites the surface mode~\cite{PhysRevLett.83.2498,PhysRevLett.84.806,science.1060182,PhysRevLett.92.050403,PhysRevA.91.013603,Zwierlein2005,Kwon2021,Poli2025}. In contrast, our method begins with macroscopic orbital angular momentum injection. Then, as the wave packet extends over the whole round, different parts of the matter wave will overlap and coherently interfere. The vortex arrays are formed from the nonzero flow velocity circulation \(I=\int_C\bm{v}\cdot\text{d}\bm{l}\), such as the integration over the yellow loop in the bottom left panel of Fig.~\ref{fig4}(a), in which vortices act like ball bearings.

The time-linear dependency of arc distances $d_{\rm v}$ between vortices is a signature of the vortex formation mechanism. First, flow velocity circulation \(I\) with fixed azimuthal span is roughly proportional to the radius distance between orbits (i.e., the radius width of the integration loop), which decreases inversely with time due to the azimuthal elongation. Second, $d_{\rm v}$ is inversely proportional to velocity circulation difference between orbits, as long as these vortices are all singly quantized. Therefore $d_{\rm v}$ increases linearly with time, as shown in Fig.~\ref{fig4}(e). This process is accelerated when  \(|\beta|\) increases [inset of Fig.~\ref{fig4}(e)], which is due to the greater difference in linear velocity between different radii and the faster decrease in distance between neighboring orbits.

% This evolutionary pathway represents a significant departure from traditional vortex nucleation methods. Rather than externally imposing rotation, our mechanism harnesses self-organizing principles of quantum many-body systems to achieve vortex matter through spontaneous symmetry breaking. The topological protection of angular momentum during the transition from stage three to stage four ensures precise control over the final vortex number, which equals the total angular momentum divided by reduced Planck's constant. This offers new opportunities for quantum fluid engineering beyond conventional approaches.

\section{\label{sec6}Conclusions}

In summary, we have demonstrated the emergence of quantum tidal locking in Bose-Einstein condensates undergoing orbital motion in anharmonic potentials. Through a combined theoretical, numerical, and experimental approach, we established that the rotation-orbit coupling in such systems drives the condensate into a synchronized state where intrinsic rotation locks to orbital motion---a quantum analog of classical tidal locking phenomena. Our findings reveal a key characteristic distinct from classical synchronization: the dynamic evolution toward vortex array formation.%(i) the degeneracy of rotational and orbital degrees of freedom, (ii) the phase-dependent symmetry breaking that underlies the locking mechanism, and (iii)  the dynamic evolution toward vortex array formation.

In our current experiments, the atom number per condensate is limited to about $1\times10^4$, which leads to deteriorated signal-to-noise ratio after multiple orbital cycles and eventual disappearance of the atomic wave packet. Future improvements should focus on increasing the atom number to sustain observable dynamics over longer timescales.

Looking forward, quantum tidal locking opens several promising research directions. The spontaneous symmetry breaking and rotation-orbit coupling mechanism underlying this phenomenon provides fertile ground for exploring the decoherence of COM degrees of freedom, nonequilibrium quantum dynamics, and pattern formation in confined geometries. The predicted formation of ring-shaped vortex arrays invites experimental investigation with longer BEC lifetimes, tunable interaction strength, and alternative trapping anharmonicity. Furthermore, this mechanism could be extended to explore topological phase transitions in other quantum fluid systems, such as fermionic superfluids or quantum droplets. 

\section*{acknowledgments}
This work was supported by the National Key Research and Development Program of China (Grant No. 2021YFA0718300 and No. 2021YFA1400900), the National Natural Science Foundation of China (Grant No. 92365208), and the Space Application System of China Manned Space Program. We thank Andreas Hemmerich for enlightening discussions.
%%%%%%%%%%%%%%%%%%%%%%%%%%%%%%%%%%%%%%%%%%%%%%%%%%%%%%%%%%%%%%%%%%%%%%

\section*{Data Availability}
The experimental and theoretical data presented in the main figures are available for download from an open repository~\cite{zenodo}. The code used for the numerical simulation in this paper is openly available~\cite{myrepo}.

% \section*{End Matter}
\section*{\label{AppxA}Appendix A: Magnitude relation of \(\omega_r\), \(\omega_\theta\), and \(\Omega\)}
The local trapping frequencies at \(\bm{X}\) are
\begin{align} 
    \label{eqA1}
    \omega_r&= \omega_0\sqrt{(1-2\beta)+\frac{6\beta}{r_0^2}|\bm{X}|^2}, \tag{A1}\\
    \label{eqA2}
    \omega_\theta &= \omega_0\sqrt{(1-2\beta)+\frac{2\beta}{r_0^2}|\bm{X}|^2}. \tag{A2}
\end{align}

Imaging two mass points in the anharmonic trap [Eq.~\eqref{eq1}], doing circular motion, each at the radius \(r_c\pm\Delta r\) (set \(r_c = |\bm{X}|\)). The average centripetal force of them is
\begin{equation}
\label{eqA3}
\bar{F} = m\omega_0^2\left((1-2\beta)r_c+\frac{2\beta}{r_0}(r_c^3+3r_c\Delta r^2)\right),  \tag{A3}
\end{equation}
which is in between \(m\omega_r^2r_c\) and \(m\omega_\theta^2r_c\), under the condition \(\Delta r/r_c\ll1\). For a wave packet, the centripetal force of COM \(\frac{\text{d}}{\text{d}t}\braket{\bm{r}} = -\braket{\bm{\nabla} V}\) is the sum of \(\bar{F}\) of a series of mass point pairs. Thus, \(\Omega\) is between \(\omega_r\) and \(\omega_\theta\).

\section*{\label{AppxB}Appendix B: Noncommutative space}
Transform \((\hat r,\hat r_\theta;\hat p_r,\hat p_\theta)\) into a new set of phase space
\begin{align*}
\label{eqA4}
\hat\xi = \frac{1}{2}\left[\hat r-\frac{\hat p_\theta}{m\omega_r}\right],
\hat\eta = \frac{1}{2}\left[\hat r_\theta-\frac{\hat p_r}{m\omega_\theta}\right],\\
\hat X = \frac{1}{2}\left[\hat r+\frac{\hat p_\theta}{m\omega_r}\right],
\hat Y = \frac{1}{2}\left[\hat r_\theta+\frac{\hat p_r}{m\omega_\theta}\right].\tag{A4}
\end{align*}
(We use \(\hat R\) and \(\hat R_\theta\) in the main text to avoid confusing \(\hat X\) with COM motion \(\bm{X}\)). Here \(\hat \xi\) and \(\hat \eta\), \(\hat X\) and \(\hat Y\) are not commutable
\[
\label{eqA5}
[\hat \xi,\hat \eta] = -[\hat X, \hat Y] = \frac{i\hbar}{2m\sqrt{\omega_r\omega_\theta}}.\tag{A5}
\]

According to \([\hat A,f(\hat B)]=[\hat A,\hat B]f'(\hat B)\), where \(f(\hat B)\) is an analytic function of \(\hat B\) and \([[\hat A, \hat B], \hat{B}] = 0\), we can derive the two components of the drift velocity
\begin{align*}
\label{eqA6}
\frac{\text{d}}{\text{d}t}\hat X &= \frac{1}{i\hbar}[\hat X,\hat V_\text{gc}] = -m\sqrt{\omega_r\omega_\theta}\frac{\partial}{\partial\hat Y}\hat V_\text{gc},\\
\frac{\text{d}}{\text{d}t}\hat Y &= \frac{1}{i\hbar}[\hat Y,\hat V_\text{gc}] = m\sqrt{\omega_r\omega_\theta}\frac{\partial}{\partial\hat X}\hat V_\text{gc},\tag{A6}
\end{align*}
thus \(\bm{v}_d = \bm{\Omega}\times\bm{\nabla} V_\text{gc}/2m\Omega\sqrt{\omega_r\omega_\theta}\).

\section*{\label{AppxC}Appendix C: Interaction strength \(g\) in simulation}
We assumed that the degree of freedom in \(z\) direction is locked in the ground state of \(V(z)=\frac{1}{2}m\omega_z^2z^2\). The wavefunction \(\psi(x,y,z)=\psi_0(x,y)\psi_1(z)\), where
\[
\label{eqA7}
\psi_1(z)=\left(\frac{m\omega_z}{\pi\hbar}\right)^{1/4}e^{-\frac{m\omega_z}{2\hbar}z^2}. \tag{A7}
\]
However, in numerical simulation, we only simulate \(\psi_0(x,y)\) and exploit a 2D grid to represent it, due to limited computing resources. Thus the wavefunction in the simulation is
\[
\label{eqA8}
\psi'_1(z) = 
\begin{cases}
1/\sqrt{x_0}, & 0 < z < x_0 \\
0, & \text{otherwise}
\end{cases},\tag{A8}
\]
where \(x_0\) is the dimensionless length unit of simulation (set to \(1\upmu\text{m}\)). As for the interaction strength \(g\), we can use neither the commonly used expression \(g=4\pi \hbar^2a_s/m\) in Ref.~\cite{RevModPhys.71.463}, since the wavefunction is incorrect, nor the interaction strength of 2D Gross-Pitaevskii equation~\cite{kim1999, Winkler2003, Lieb2001, PhysRevA.109.043304, PhysRevLett.126.244101}, since we are simulating a three-dimensional BEC. We need to renormalize \(g\) by equating the interacting term in the energy functional
\[
\label{eqA9}\int\text{d}^3\bm{r}\frac{g}{2}|\psi_0(x,y)\psi_1(z)|^4 = 
\int\text{d}^3\bm{r}\frac{g'}{2}|\psi_0(x,y)\psi'_1(z)|^4,\tag{A9}
\]
and derive the interaction strength in simulation
\[
\label{eqA10}
g' = \sqrt{\frac{m\omega_z x_0^2}{2\pi\hbar}}\frac{4\pi\hbar^2a_s}{m}.\tag{A10}
\]
In a quadrupole trap, BEC generally exhibits a more dispersed distribution along \(z\), so we set \(\omega_z = \omega_0/2\).
\section*{\label{AppxD}Appendix D: Another way to impose initial rotation}

\begin{figure}[t!]
	\includegraphics[width=\linewidth]{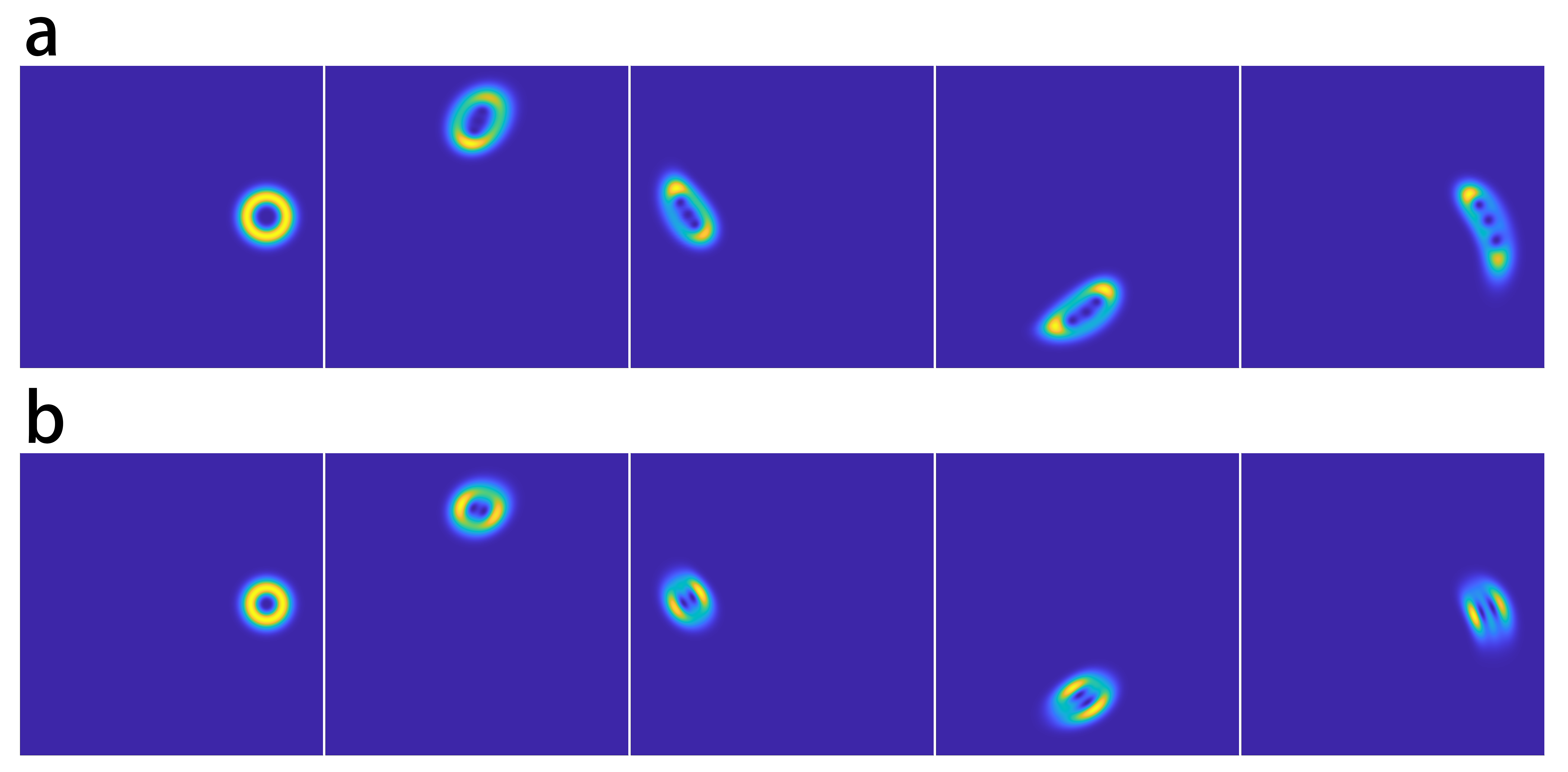} % Here is how to import EPS art
	\caption{\label{fig6} \(r_0=|\bm{X}|=80\upmu\text{m}\), \(\beta = -0.1\), \(\omega_0/2\pi = 1.25\text{ Hz}\), and \(g = 0\). (a) The initial state is \(\ket{l=3, m=3}\). (b) The initial state is \(\ket{l = 2, m = -2}.\)}
\end{figure}

In Fig.~\ref{fig23}(c), the initial angular momentum was imposed by quadrupole velocity field, similar with Fig.~\ref{fig23}(b). Here we imposed the initial angular momentum by making the initial state an eigenstate of \(\hat{\bm L} ^2\) and \(\hat{L}_z\), and discovered that the numerical result [Fig.~\ref{fig6}] is highly consistent with Fig.~3 in Ref.~\cite{CRPHYS_2023__24_S3_241_0}.

\nocite{*}

\bibliography{apssamp}% Produces the bibliography via BibTeX.

@article{RevModPhys.82.3045,
  title = {Colloquium: Topological insulators},
  author = {Hasan, M. Z. and Kane, C. L.},
  journal = {Rev. Mod. Phys.},
  volume = {82},
  issue = {4},
  pages = {3045--3067},
  numpages = {0},
  year = {2010},
  month = {Nov},
  publisher = {American Physical Society},
  doi = {10.1103/RevModPhys.82.3045},
  url = {https://link.aps.org/doi/10.1103/RevModPhys.82.3045}
}

@article{Galitski2013,
   author = {Galitski, Victor and Spielman, Ian B.},
   title = {Spin-orbit coupling in quantum gases},
   journal = {Nature},
   volume = {494},
   number = {7435},
   pages = {49-54},
   ISSN = {1476-4687},
   DOI = {10.1038/nature11841},
   url = {https://doi.org/10.1038/nature11841},
   year = {2013},
   type = {Journal Article}
}

@article{Zhai_2015,
doi = {10.1088/0034-4885/78/2/026001},
url = {https://doi.org/10.1088/0034-4885/78/2/026001},
year = {2015},
month = {feb},
publisher = {IOP Publishing},
volume = {78},
number = {2},
pages = {026001},
author = {Zhai, Hui},
title = {Degenerate quantum gases with spin-orbit coupling: a review},
journal = {Reports on Progress in Physics}
}

@article{Winkler2003,
author = {Winkler, Roland},
year = {2003},
month = {01},
pages = {},
title = {Spin-Orbit Coupling Effects in Two-Dimensional Electron and Hole Systems},
volume = {191},
isbn = {978-3-540-01187-3},
journal = {Springer Tracts in Modern Physics},
doi = {10.1007/b13586}
}

@article{Goldreich1966,
  title={History of the lunar orbit},
  author={Goldreich, Peter},
  journal={Rev. Geophys.},
  volume={4},
  number={4},
  pages={411--439},
  year={1966},
  publisher={Wiley Online Library}
}

@article{Williams2014,
  title={Lunar rotational dissipation in solid body and molten core},
  author={Williams, James G and Boggs, Dale H},
  journal={Geophys. Res. Planets},
  volume={119},
  number={7},
  pages={1546--1578},
  year={2014},
  publisher={Wiley Online Library}
}

@article{Cooper2008,
author = {N.R. Cooper},
title = {Rapidly rotating atomic gases},
journal = {Adv. Phys.},
volume = {57},
number = {6},
pages = {539--616},
year = {2008},
publisher = {Taylor \& Francis},
doi = {10.1080/00018730802564122},
URL = {https://doi.org/10.1080/00018730802564122}
}

@article{doi:10.1126/science.aba7202,
author = {Richard J. Fletcher  and Airlia Shaffer  and Cedric C. Wilson  and Parth B. Patel  and Zhenjie Yan  and Valentin Crépel  and Biswaroop Mukherjee  and Martin W. Zwierlein },
title = {Geometric squeezing into the lowest {Landau} level},
journal = {Science},
volume = {372},
number = {6548},
pages = {1318-1322},
year = {2021},
doi = {10.1126/science.aba7202}
}

@article{PhysRevA.105.023310,
  title = {Rotating {Bose} gas dynamically entering the lowest {Landau} level},
  author = {Sharma, Vaibhav and Mueller, Erich J.},
  journal = {Phys. Rev. A},
  volume = {105},
  issue = {2},
  pages = {023310},
  numpages = {7},
  year = {2022},
  month = {Feb},
  publisher = {American Physical Society},
  doi = {10.1103/PhysRevA.105.023310},
  url = {https://link.aps.org/doi/10.1103/PhysRevA.105.023310}
}

@article{CRPHYS_2023__24_S3_241_0,
     author = {Valentin Cr\'epel and Ruixiao Yao and Biswaroop Mukherjee and Richard Fletcher and Martin Zwierlein },
     title = {Geometric squeezing of rotating quantum gases into the lowest {Landau} level},
     journal = {C. R. Phys.},
     pages = {241--262},
     publisher = {Acad\'emie des sciences, Paris},
     volume = {24},
     number = {S3},
     year = {2023},
     doi = {10.5802/crphys.173},
}

@article{chen2025,
  title = {Dynamical generation of geometric squeezing in interacting {Bose-Einstein} condensates},
  author = {Chen, Li and Zhu, Fei and Tang, Zheng and Zeng, Liang and Lee, Jae Joon and Pu, Han},
  journal = {Phys. Rev. A},
  volume = {112},
  issue = {4},
  pages = {043319},
  numpages = {9},
  year = {2025},
  publisher = {American Physical Society},
  doi = {10.1103/sf74-lk9d},
  url = {https://link.aps.org/doi/10.1103/sf74-lk9d}
}

@article{PhysRevLett.86.377,
  title = {Overcritical Rotation of a Trapped {Bose-Einstein} Condensate},
  author = {Recati, A. and Zambelli, F. and Stringari, S.},
  journal = {Phys. Rev. Lett.},
  volume = {86},
  issue = {3},
  pages = {377--380},
  numpages = {0},
  year = {2001},
  publisher = {American Physical Society},
  doi = {10.1103/PhysRevLett.86.377},
  url = {https://link.aps.org/doi/10.1103/PhysRevLett.86.377}
}

@dataset{zenodo,
	title = {Data for ``{Quantum} Tidal Locking in Orbiting {Bose-Einstein} Condensates''},
	author = {Fan, Yaoyuan and Shi, Shuoyu and Cao, Lang and He, Ziyue and Zhang, Qiuxin and Hu, Dong and Wang, Yu and Wang, Qing and Zhou, Tianwei and Zhou, Xiaoji},
	month = apr,
    year = 2026,
    publisher = {Zenodo},
	  doi = {https://doi.org/10.5281/zenodo.19875235},
    url = {https://doi.org/10.5281/zenodo.19875235}
}

@misc{myrepo,
  note={{C}ode available from \url{https://github.com/ShiShuoyu/BEC_Dynamics_2D/releases/tag/v1.2.3}},
}

@article{doi:10.1137/0723033,
author = {Weideman, J. A. C. and Herbst, B. M.},
title = {Split-Step Methods for the Solution of the Nonlinear {Schrödinger} Equation},
journal = {SIAM J. Numer. Anal.},
volume = {23},
number = {3},
pages = {485-507},
year = {1986},
doi = {10.1137/0723033},
}

@article{BAO2003318,
title = {Numerical solution of the {Gross-Pitaevskii} equation for {Bose-Einstein} condensation},
journal = {J. Comput. Phys.},
volume = {187},
number = {1},
pages = {318-342},
year = {2003},
issn = {0021-9991},
doi = {https://doi.org/10.1016/S0021-9991(03)00102-5},
url = {https://www.sciencedirect.com/science/article/pii/S0021999103001025},
author = {Weizhu Bao and Dieter Jaksch and Peter A. Markowich},
}

@article{WANG200517,
title = {Numerical studies on the split-step finite difference method for nonlinear {Schrödinger} equations},
journal = {Appl. Math. Comput.},
volume = {170},
number = {1},
pages = {17-35},
year = {2005},
issn = {0096-3003},
doi = {https://doi.org/10.1016/j.amc.2004.10.066},
url = {https://www.sciencedirect.com/science/article/pii/S0096300304008689},
author = {Hanquan Wang},
}

@article{RevModPhys.71.463,
  title = {Theory of {Bose-Einstein} condensation in trapped gases},
  author = {Dalfovo, Franco and Giorgini, Stefano and Pitaevskii, Lev P. and Stringari, Sandro},
  journal = {Rev. Mod. Phys.},
  volume = {71},
  issue = {3},
  pages = {463--512},
  numpages = {0},
  year = {1999},
  month = {Apr},
  publisher = {American Physical Society},
  doi = {10.1103/RevModPhys.71.463},
  url = {https://link.aps.org/doi/10.1103/RevModPhys.71.463}
}

@misc{kim1999,
      title={Two-Dimensional Gross-Pitaevskii Equation: Theory of {Bose-Einstein} Condensation and the Vortex State}, 
      author={Sang-Hoon Kim and Changyeon Won and Sung Dahm Oh and Wonho Jhe},
      year={1999},
      eprint={cond-mat/9904087},
      archivePrefix={arXiv},
      primaryClass={cond-mat},
      url={https://arxiv.org/abs/cond-mat/9904087}, 
}

@article{Lieb2001,
      title={A Rigorous Derivation of the {Gross-Pitaevskii} Energy Functional for a Two-dimensional {Bose} Gas}, 
      author={Elliott H. Lieb and Robert Seiringer and Jakob Yngvason},
      year={2001},
      journal = {Commun. Math. Phys.},
      volume = {224},
      pages = {17--31},
      url={https://doi.org/10.1007/s002200100533}, 
}

@article{PhysRevLett.126.244101,
  title = {Rotating Multidimensional Quantum Droplets},
  author = {Dong, Liangwei and Kartashov, Yaroslav V.},
  journal = {Phys. Rev. Lett.},
  volume = {126},
  issue = {24},
  pages = {244101},
  numpages = {7},
  year = {2021},
  month = {Jun},
  publisher = {American Physical Society},
  doi = {10.1103/PhysRevLett.126.244101},
  url = {https://link.aps.org/doi/10.1103/PhysRevLett.126.244101}
}

@article{Yaoyuan,
title = {Spin polarization control via magnetic field in dissipative bosonic systems},
journal = {Front. Phys.},
volume = {21},
pages = {012200-},
year = {2026},
issn = {2095-0462},
doi = {https://doi.org/10.15302/frontphys.2026.012200},
url = {https://journal.hep.com.cn/fop/EN/10.15302/frontphys.2026.012200},
author = {Yaoyuan Fan and Shuoyu Shi and Lang Cao and Qiuxin Zhang and Dong Hu and Yu Wang and Xiaoji Zhou},
}

@article{PhysRevA.102.063330,
  title = {Breakup of rotating asymmetric quartic-quadratic trapped condensates},
  author = {Brito, Leonardo and Andriati, Alex and Tomio, Lauro and Gammal, Arnaldo},
  journal = {Phys. Rev. A},
  volume = {102},
  issue = {6},
  pages = {063330},
  numpages = {13},
  year = {2020},
  month = {Dec},
  publisher = {American Physical Society},
  doi = {10.1103/PhysRevA.102.063330},
  url = {https://link.aps.org/doi/10.1103/PhysRevA.102.063330}
}

@article{PhysRevA.109.043304,
  title = {Rotating quantum droplets confined in an anharmonic potential},
  author = {Nikolaou, S. and Kavoulakis, G. M. and \"Ogren, M.},
  journal = {Phys. Rev. A},
  volume = {109},
  issue = {4},
  pages = {043304},
  numpages = {11},
  year = {2024},
  month = {Apr},
  publisher = {American Physical Society},
  doi = {10.1103/PhysRevA.109.043304},
  url = {https://link.aps.org/doi/10.1103/PhysRevA.109.043304}
}

@article{Kavoulakis_2003,
url = {https://doi.org/10.1088/1367-2630/5/1/351},
year = {2003},
publisher = {},
volume = {5},
number = {1},
pages = {51},
author = {Kavoulakis, G M and Baym, Gordon},
title = {Rapidly rotating {Bose-Einstein} condensates in
anharmonic potentials},
journal = {New J. Phys.}
}

@article{PhysRevA.71.013605,
  title = {Rapid rotation of a {Bose-Einstein} condensate in a harmonic plus quartic trap},
  author = {Fetter, Alexander L. and Jackson, B. and Stringari, S.},
  journal = {Phys. Rev. A},
  volume = {71},
  issue = {1},
  pages = {013605},
  numpages = {9},
  year = {2005},
  month = {Jan},
  publisher = {American Physical Society},
  doi = {10.1103/PhysRevA.71.013605},
  url = {https://link.aps.org/doi/10.1103/PhysRevA.71.013605}
}

@article{PhysRevLett.83.2498,
  title = {Vortices in a {Bose-Einstein} Condensate},
  author = {Matthews, M. R. and Anderson, B. P. and Haljan, P. C. and Hall, D. S. and Wieman, C. E. and Cornell, E. A.},
  journal = {Phys. Rev. Lett.},
  volume = {83},
  issue = {13},
  pages = {2498--2501},
  year = {1999},
  publisher = {American Physical Society},
  doi = {10.1103/PhysRevLett.83.2498},
  url = {https://link.aps.org/doi/10.1103/PhysRevLett.83.2498}
}

@article{PhysRevLett.84.806,
  title = {Vortex Formation in a Stirred {Bose-Einstein} Condensate},
  author = {Madison, K. W. and Chevy, F. and Wohlleben, W. and Dalibard, J.},
  journal = {Phys. Rev. Lett.},
  volume = {84},
  issue = {5},
  pages = {806--809},
  numpages = {0},
  year = {2000},
  month = {Jan},
  publisher = {American Physical Society},
  doi = {10.1103/PhysRevLett.84.806},
  url = {https://link.aps.org/doi/10.1103/PhysRevLett.84.806}
}

@article{PhysRevLett.84.2056,
  title = {Observation of the Scissors Mode and Evidence for Superfluidity of a Trapped {Bose-Einstein} Condensed Gas},
  author = {Marag\`o, O. M. and Hopkins, S. A. and Arlt, J. and Hodby, E. and Hechenblaikner, G. and Foot, C. J.},
  journal = {Phys. Rev. Lett.},
  volume = {84},
  issue = {10},
  pages = {2056--2059},
  numpages = {0},
  year = {2000},
  month = {Mar},
  publisher = {American Physical Society},
  doi = {10.1103/PhysRevLett.84.2056},
  url = {https://link.aps.org/doi/10.1103/PhysRevLett.84.2056}
}

@article{
science.1060182,
author = {J. R. Abo-Shaeer  and C. Raman  and J. M. Vogels  and W. Ketterle },
title = {Observation of Vortex Lattices in {Bose-Einstein} Condensates},
journal = {Science},
volume = {292},
number = {5516},
pages = {476-479},
year = {2001},
doi = {10.1126/science.1060182},
URL = {https://www.science.org/doi/abs/10.1126/science.1060182}
}

@article{PhysRevLett.92.040404,
  title = {Rapidly Rotating {Bose-Einstein} Condensates in and near the Lowest {Landau} Level},
  author = {Schweikhard, V. and Coddington, I. and Engels, P. and Mogendorff, V. P. and Cornell, E. A.},
  journal = {Phys. Rev. Lett.},
  volume = {92},
  issue = {4},
  pages = {040404},
  numpages = {4},
  year = {2004},
  month = {Jan},
  publisher = {American Physical Society},
  doi = {10.1103/PhysRevLett.92.040404},
  url = {https://link.aps.org/doi/10.1103/PhysRevLett.92.040404}
}

@article{PhysRevLett.92.050403,
  title = {Fast Rotation of a {Bose-Einstein} Condensate},
  author = {Bretin, Vincent and Stock, Sabine and Seurin, Yannick and Dalibard, Jean},
  journal = {Phys. Rev. Lett.},
  volume = {92},
  issue = {5},
  pages = {050403},
  numpages = {4},
  year = {2004},
  month = {Feb},
  publisher = {American Physical Society},
  doi = {10.1103/PhysRevLett.92.050403},
  url = {https://link.aps.org/doi/10.1103/PhysRevLett.92.050403}
}

@article{RevModPhys.81.647,
  title = {Rotating trapped {Bose-Einstein} condensates},
  author = {Fetter, Alexander L.},
  journal = {Rev. Mod. Phys.},
  volume = {81},
  issue = {2},
  pages = {647--691},
  numpages = {0},
  year = {2009},
  month = {May},
  publisher = {American Physical Society},
  doi = {10.1103/RevModPhys.81.647},
  url = {https://link.aps.org/doi/10.1103/RevModPhys.81.647}
}

@article{PhysRevA.91.013603,
  title = {Rotating a {Bose-Einstein} condensate by shaking an anharmonic axisymmetric magnetic potential},
  author = {Kang, Seji and Choi, J. and Seo, S. W. and Kwon, W. J. and Shin, Y.},
  journal = {Phys. Rev. A},
  volume = {91},
  issue = {1},
  pages = {013603},
  numpages = {8},
  year = {2015},
  publisher = {American Physical Society},
  doi = {10.1103/PhysRevA.91.013603}
}

@article{Mukherjee2022,
   author = {Mukherjee, Biswaroop and Shaffer, Airlia and Patel, Parth B. and Yan, Zhenjie and Wilson, Cedric C. and Crépel, Valentin and Fletcher, Richard J. and Zwierlein, Martin},
   title = {Crystallization of bosonic quantum {Hall} states in a rotating quantum gas},
   journal = {Nature},
   volume = {601},
   number = {7891},
   pages = {58-62},
   ISSN = {1476-4687},
   DOI = {10.1038/s41586-021-04170-2},
   url = {https://doi.org/10.1038/s41586-021-04170-2},
   year = {2022},
   type = {Journal Article}
}

@article{Zwierlein2005,
   author = {Zwierlein, M. W. and Abo-Shaeer, J. R. and Schirotzek, A. and Schunck, C. H. and Ketterle, W.},
   title = {Vortices and superfluidity in a strongly interacting {Fermi} gas},
   journal = {Nature},
   volume = {435},
   number = {7045},
   pages = {1047-1051},
   ISSN = {1476-4687},
   DOI = {10.1038/nature03858},
   url = {https://doi.org/10.1038/nature03858},
   year = {2005},
   type = {Journal Article}
}

@article{Kwon2021,
   author = {Kwon, W. J. and Del Pace, G. and Xhani, K. and Galantucci, L. and Muzi Falconi, A. and Inguscio, M. and Scazza, F. and Roati, G.},
   title = {Sound emission and annihilations in a programmable quantum vortex collider},
   journal = {Nature},
   volume = {600},
   number = {7887},
   pages = {64-69},
   ISSN = {1476-4687},
   DOI = {10.1038/s41586-021-04047-4},
   url = {https://doi.org/10.1038/s41586-021-04047-4},
   year = {2021},
   type = {Journal Article}
}

@article{Lunt2024,
	title        = {Realization of a {Laughlin} State of Two Rapidly Rotating Fermions},
	author       = {Lunt, Philipp and Hill, Paul and Reiter, Johannes and Preiss, Philipp M. and Ga\l{}ka, Maciej and Jochim, Selim},
	year         = {2024},
	journal      = {Phys. Rev. Lett.},
	publisher    = {APS},
	volume       = {133},
	pages        = {253401},
	doi          = {10.1103/PhysRevLett.133.253401}
}

@article{Poli2025,
   author = {Poli, Elena and Litvinov, Andrea and Casotti, Eva and Ulm, Clemens and Klaus, Lauritz and Mark, Manfred J. and Lamporesi, Giacomo and Bland, Thomas and Ferlaino, Francesca},
   title = {Synchronization in rotating supersolids},
   journal = {Nature Phys.},
   ISSN = {1745-2481},
   DOI = {10.1038/s41567-025-03065-7},
   url = {https://doi.org/10.1038/s41567-025-03065-7},
   year = {2025},
   type = {Journal Article}
}

% ****** End of file apssamp.tex ******
\end{document}